\documentclass[aps,prb,twocolumn,floatfix,superscriptaddress]{revtex4-2}
\usepackage{amsmath}
\usepackage{amssymb}
\usepackage{multirow}
\usepackage{graphicx}
\usepackage{babel}
\usepackage{hyperref}
\usepackage{natbib}
\usepackage{siunitx}
\usepackage[T1]{fontenc}
\usepackage{color}
\usepackage{epstopdf}
\usepackage{siunitx}

\begin{document}

\title{Tunneling magnetoresistance in MgO tunnel junctions with Fe-based leads in empirically corrected density functional theory}
\date{\today}

\author{G. G. Baez Flores}
\affiliation{Department of Physics and Astronomy and Nebraska Center for Materials and Nanoscience, University of Nebraska-Lincoln, Lincoln, Nebraska, 68588, USA}

\author{M. van Schilfgaarde}
\affiliation{National Renewable Energy Laboratory, Golden, Colorado 80401, USA}

\author{K. D. Belashchenko}
\affiliation{Department of Physics and Astronomy and Nebraska Center for Materials and Nanoscience, University of Nebraska-Lincoln, Lincoln, Nebraska, 68588, USA}

\begin{abstract}
    The minority-spin Fe/MgO interface states are at the Fermi level in density functional theory (DFT), but experimental evidence and GW calculations place them slightly higher in energy. This small shift can strongly influence tunneling magnetoresistance (TMR) in junctions with a thin MgO barrier and its dependence on the concentration of Co in the electrodes. Here, an empirical potential correction to DFT is introduced to shift the interface states up to match the tunnel spectroscopy data. With this shift, TMR in Fe/MgO/Fe junctions exceeds 800\% and 3000\% at 3 and 4 monolayers (ML) of MgO, respectively. We further consider the effect of alloying of the Fe electrodes with up to 30\% Co or 10\% V, treating them in the coherent potential approximation (CPA). Alloying with Co broadens the interface states and brings a large incoherent minority-spin spectral weight to the Fermi level. Alloying with V brings the minority-spin resonant states close to the Fermi level. However, in both cases the minority-spin spectral weight at the Fermi level resides primarily at the periphery of the Brillouin zone, which is favorable for spin filtering. Using convolutions of $\mathbf{k}_\parallel$-resolved barrier densities of states calculated in CPA, it is found that TMR is strongly reduced by alloying with Co or V but still remains above 500\% at 4 ML of MgO up to 30\% of Co or 5\% V. At 5 ML, the TMR increases above 1000\% in all systems considered. On the other hand, while TMR declines sharply with increasing bias up to 0.2 eV in the MTJ with pure Fe leads, it remains almost constant up to 0.5 eV if leads are alloyed with Co.
\end{abstract}

\maketitle

\section{Introduction}

Epitaxial Fe/MgO/Fe(001) magnetic tunnel junctions (MTJ) exhibit large tunneling magnetoresistance (TMR) \cite{Butler,Yuasa2004,Parkin2004,Yuasa-handbook} thanks to the symmetry-enforced spin filtering \cite{MacLaren1999,Mavropoulos2000} facilitated by the matching of the majority-spin $\Delta_1$ band in Fe to the evanescent state in MgO with the same symmetry. In device applications, the leads usually employ a Fe-Co alloy which helps grow the MTJ stack with a pinned magnetic layer \cite{Yuasa-handbook}. Furthermore, spin transfer torque-based magnetic random-access memory (MRAM) requires MTJs with a low resistance-area product ($RA$), which attracts interest to the properties of MTJs with an ultrathin MgO layer \cite{Yakushiji_2010}.

The Fe(001) surface carries a surface resonant band near the Fermi level for minority spins \cite{Stroscio1995}. This band is also present at the Fe(001)/MgO interface and can contribute significantly to the tunneling current in the antiparallel (AP) configuration, reducing the TMR \cite{Butler,Tiusan2004,Tiusan2006,KBMtj,TMR-theory}. This contribution is expected to be more pronounced in ultrathin MTJs where symmetry-enforced spin filtering is less effective.

In density-functional calculations, the Fe(001)/MgO interface resonant states appear right at the Fermi level \cite{KBMtj}, but experimental tunnel spectroscopy measurements \cite{Tiusan2006,Zermatten2008} and GW calculations \cite{MTJGW} place them 0.1-0.15 eV higher in energy. Photoemission spectroscopy measurements \cite{Bonell} also show that the interface states are empty at the Fe/MgO interface and begin to be filled in Fe$_{1-x}$Co$_x$/MgO only at $x\approx0.37$. Thus, it is expected that density-functional calculations of spin-dependent tunneling in Fe(001)-based MTJs with a thin MgO barrier should strongly overestimate the conductance in the antiparallel configuration. 

In this paper, we introduce an empirical correction to bring the electronic structure of the Fe/MgO interface in better agreement with GW calculations and experimental data. Using the coherent potential approximation (CPA), we then study the electronic structure of the interfaces and TMR in FeX/MgO/FeX MTJs, where X stands for alloying with Co or V. Alloying with V is considered because it was reported to reduce the density of dislocations \cite{Bonell2010}.
Using the approximate factorization of the transmission probability \cite{Belashchenko2004,TMR-theory}, we use convolutions of the barrier densities of states (DOS) to study the influence of alloying on the TMR, including its bias dependence. 

The paper is organized as follows. In Section \ref{sec:GW} we revisit the electronic structure of the Fe/MgO interface within the quasiparticle self-consistent $GW$ (QS$GW$) method \cite{QUESTAAL} and confirm the main conclusions of Ref. \onlinecite{MTJGW}. In Section \ref{sec:correction} we introduce an empirical correction for the Fe/MgO interface resonant states. Section \ref{sec:femgofe} deals with spin-dependent tunneling in pure Fe/MgO/Fe MTJs, and Section \ref{sec:TDOS} is focused on FeX/MgO/FeX systems. The bias dependence of TMR is addressed in Section \ref{sec:bias}. Section \ref{sec:conclusions} concludes the paper.

\section{Sensitivity of the F\lowercase{e}/M\lowercase{g}O surface states to lattice and magnetic structure}
\label{sec:GW}

In this section we study the electronic structure of the Fe/MgO interface using the QS\emph{GW} code in Questaal \cite{QUESTAAL}. Our calculations are similar to those in Ref.~\cite{MTJGW} but include a larger basis set and more careful checks of convergence in the \emph{k} mesh and the number of interfacial layers.

One important effect of alloying Fe with another transition metal element is the modification of the exchange-correlation fields near the surface, which affects the alignment of the minority Fe \emph{d} surface states relative to the Fermi level $E_{F}$. Therefore, we consider several kinds of variations in both magnetic and lattice structure to probe how these surface states depends on these changes.

The first computational setup (Setup I), which is also used in the following DFT calculations, was obtained by full optimization of the Fe/MgO($N$)/Fe MTJ using the plane-wave pseudopotential VASP code \cite{VASP1,VASP2,VASP3} with the PBE exchange-correlation functional \cite{PBE}. The in-plane lattice constant was fixed at 2.834 \AA, which is the equilibrium bulk lattice constant of bcc Fe in PBE.

The second setup (Setup II) was prepared using a number of plausible assumptions. Experimentally, the bulk lattice constant of Fe is 2.866 \AA, and it is reasonable to expect that, away from the frontier layer, the Fe side of the junction assumes a bcc structure at its experimental bulk lattice constant, and similarly for the MgO side.  (As we show below, the position of the Fe surface states is somewhat sensitive to the Fe lattice constant.) Since bulk MgO has a lattice constant about 4\% larger than Fe, a better characterization of the interface would require a coincident site lattice with Fe and MgO lattice constants close to their respective bulk values.  This is possible but costly, as it would entail a complex interfacial structure.  Instead we consider a simple coincident site lattice
similar to the first parts of this work, but where we attempt to preserve as well as possible some key features of the
real interface.  Since the Fe surface state is more sensitive to the Fe structure than that of MgO, we repeat the superlattice calculations with the following model structure:
\begin{enumerate}
\item In-plane lattice constant of 2.866\,\AA, where MgO is compressed in the plane to match Fe;
\item Only frontier Fe, Mg, and O atoms are relaxed using the PBEsol \cite{PBEsol} functional.
\item Interplanar Fe spacings are set to ideal bcc Fe;
\item MgO is taken in the rocksalt structure, which is expanded by 8\% in the normal direction to preserve the bulk unit cell volume.
\end{enumerate}
The last of these assumptions approximately preserves the alignment of the MgO bands relative to the cubic rocksalt structure, which is the
main feature of interest for the junction.

\begin{figure}[htb]
       \includegraphics[width=0.85\columnwidth]{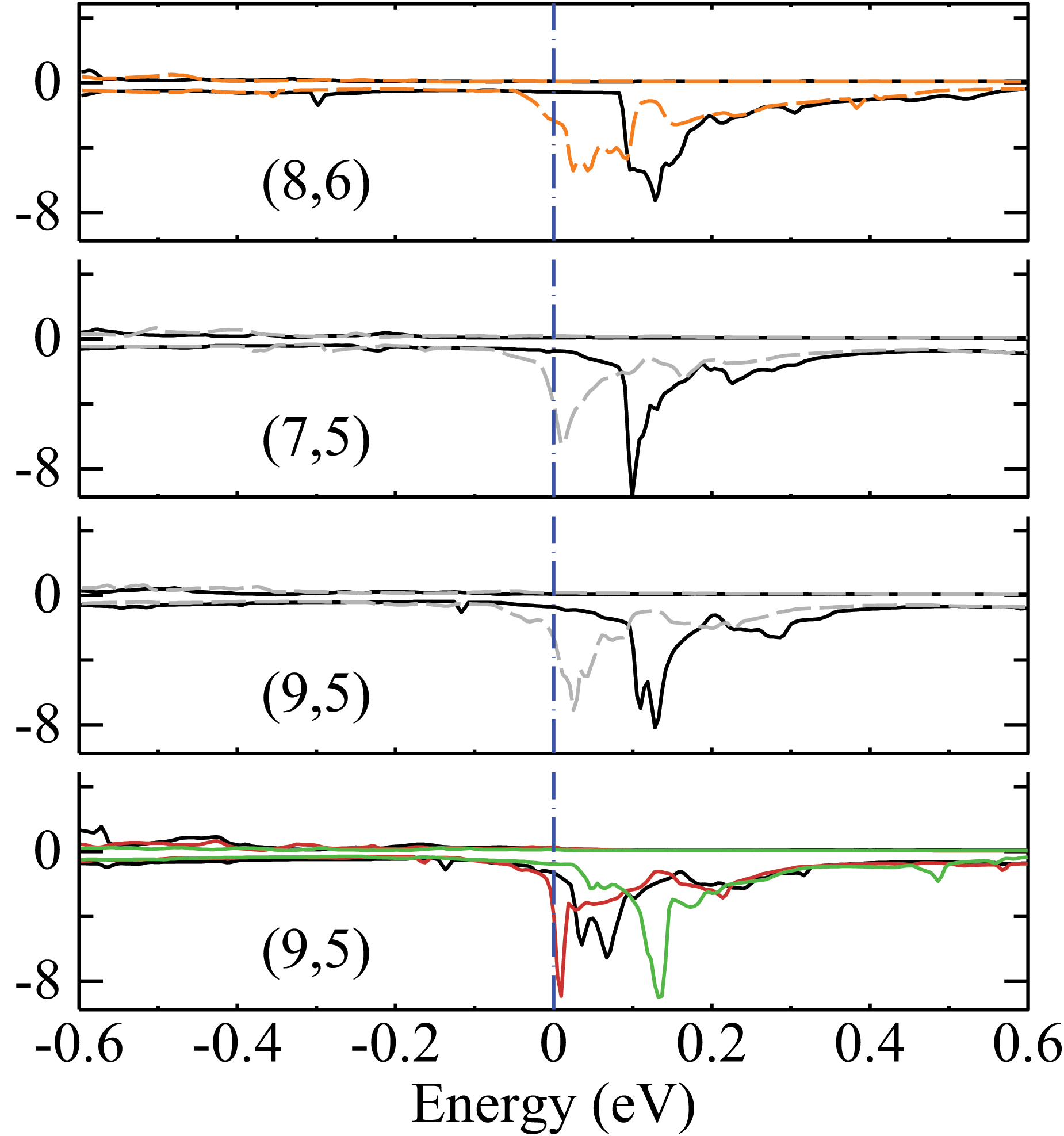}
        \caption{Density-of-states $D_f(E)$ projected onto Fe frontier atom. $D_f>0$ and $D_f<0$ correspond to
          majority and minority spin, respectively. Abscissa: energy relative to the Fermi level (eV).
          Pairs of integers ($M$,$N$) designate a multilayer with periodically repeated $M$ ML of Fe and $N$ ML of MgO.
          Orange lines for (8,6): QS\emph{GW} for Setup I. All other lines: Setup II. Dashed gray lines: DFT calculations with PBE or PBEsol functionals. Bottom panel: a (9,5) multilayer with a magnetic field of 10 mRy applied antiparallel to the magnetic moment at the frontier layer (black lines), at the adjacent Fe layer (red lines), or at the next adjacent layer (green lines).}
        \label{fig:surfaceDOS}
\end{figure}

The top panel of Fig.~\ref{fig:surfaceDOS} compares the DOS of the Fe frontier layer for Setups I and II. The \emph{d} level is shifted significantly, showing sensitivity of its position to the lattice structure.  Most of the shift can be attributed
to the smaller Fe lattice constant in the PBE-relaxed Setup I.  The middle panels show how the Fe surface state depends on the thicknesses of Fe and MgO layers in Setup II.  There is a slight dependence, but the position for (7,5), (8,6), and (9,5) structures are similar.  This demonstrates that a few monolayers are sufficient to establish the bulk properties of the
junction.  The middle panels also show the DOS calculated from the PBE functional for Setup II. QS\emph{GW} puts the minority-spin interface states 0.1 to \qty{0.15}{\electronvolt} higher than DFT, as noted above and in Ref.~\cite{MTJGW}.

\begin{table}[htb]
  \caption{Local magnetic moments $M$ ($\mu_B$) in layers 1-5 calculated for (9,5) layer MTJ.  Columns show $M$ without a field, and with an external magnetic field of 10 mRy applied antiparallel to the Fe magnetic moment.  The field is applied to a single layer: the first, second, or third layer from the interface.  The remaining layers (6-9) are equivalent to layers 1-4.}
  \begin{tabular}{|c|c|c|c|c|}
  \hline
   Layer &  $B$=0  & Layer 1 & Layer 2 & Layer 3\\
    \hline
   1   & 3.04  & 2.83    & 3.07    & 3.06 \\
   2   & 2.57  & 2.64    & 1.98    & 2.56 \\
   3   & 2.60  & 2.61    & 2.63    & 1.81 \\
   4   & 2.33  & 2.32    & 2.39    & 2.48 \\
   5   & 2.21  & 2.17    & 2.29    & 2.54 \\
    \hline
    \end{tabular}
    \label{tab:Femoment}
\end{table}

Finally, the bottom panel shows how changes in local magnetic moments affect the surface state.  An external field of 10 mRy was applied to a single Fe plane layer: the frontier layer (black), the second layer (red), or the third layer (green).  Perturbations at the first and second layers induce considerable shifts in the defect level, while
it is mostly gone for a shift induced from the third layer.  Table \ref{tab:Femoment} shows the local Fe moments $M$ on the five nonequivalent layers.  The first column shows how $M$ is strongly enhanced at the surface.  As expected, $M$ is partially quenched in the plane where $B$ is applied, but slightly enhanced in neighboring planes.  Also, the shift in the defect level is not strictly a local property of the frontier layer, but the second layer is equally important.

Clearly, variations in structure and magnetization near the Fe/MgO interface can have a notable effect on the position of the interface resonant states. On the other hand, more accurate QS\emph{GW} calculations described in this section confirm the conclusion of Ref.~\cite{MTJGW} that local exchange-correlation functionals put the minority-spin surface state 0.1 to \qty{0.15}{\electronvolt} too low. This conclusion is consistent with experimental data \cite{Tiusan2006,Zermatten2008,Bonell}.

\section{Empirical correction for the {F\lowercase{e}/M\lowercase{g}O} interface states}
\label{sec:correction}

To bring the position of the surface resonance states in better agreement with experiment, we introduce an empirical correction that can be applied to the surface Fe atoms in DFT calculations.

\begin{figure}[htb]
    \includegraphics[width=\columnwidth]{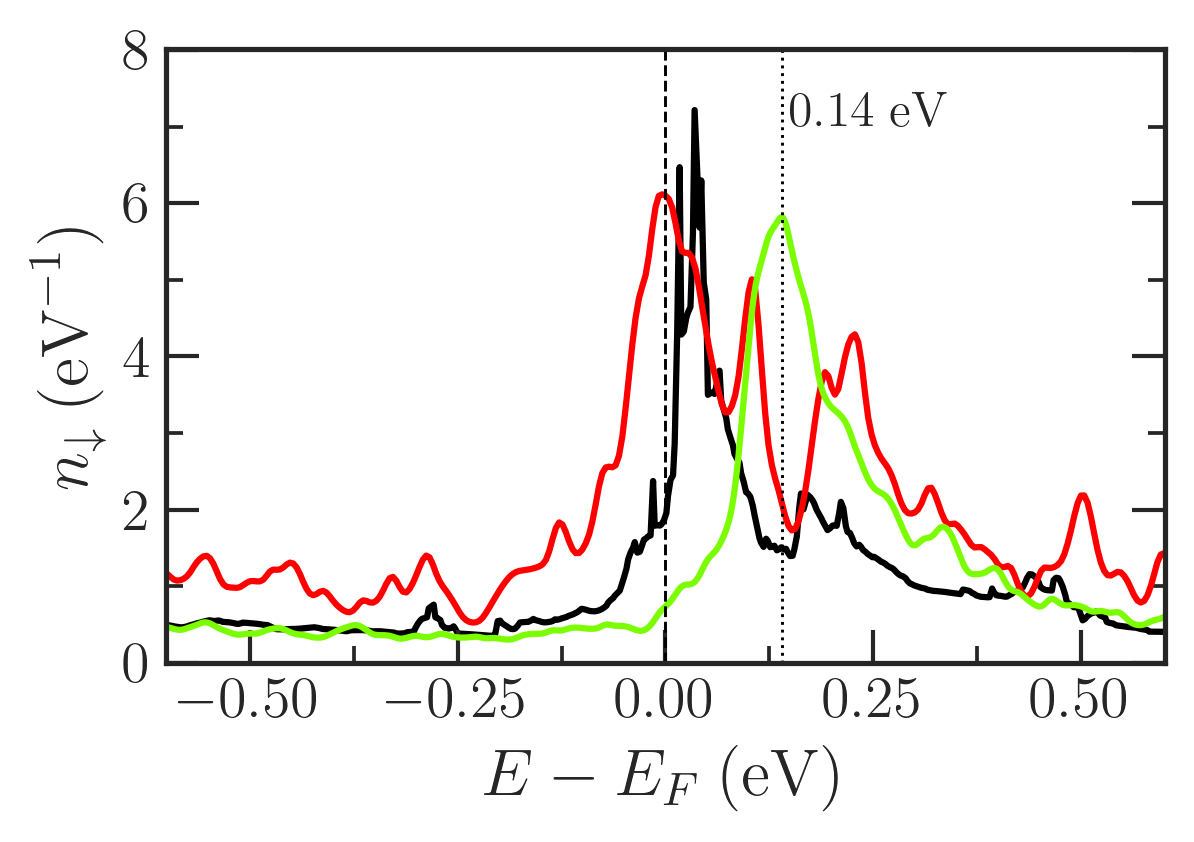}
    \caption{Minority-spin partial DOS for the interfacial Fe atom in the Fe/MgO(4)/Fe MTJ. Black: PBE calculation from VASP. Red: PBE calculation from LMTO. Green: PBE with the empirical correction from LMTO.}
    \label{fig:DOSALL}
\end{figure}

Figure \ref{fig:DOSALL}(a) shows the minority-spin DOS $n_{\downarrow}(E)$ near the Fermi level obtained using VASP (black line) and LMTO (red line) methods for the relaxed structure. The peak at the Fermi level shows the position of the interface states. The DOS from a GW calculation \cite{MTJGW} is shown in Fig. \ref{fig:DOSALL}(b). The interface states in GW are shifted up by 0.14 eV compared to DFT. We found that a similar upward shift of the interface states can be obtained by adding a 3.4 eV potential shift to the $3d$ orbitals of the Fe atoms in the surface monolayer. Figure \ref{fig:DOSALL}(a) shows the resulting minority-spin DOS (green line), which can be compared with the GW results in Fig. \ref{fig:DOSALL}(b).

\begin{figure}[ht]
    \includegraphics[width=0.95\columnwidth]{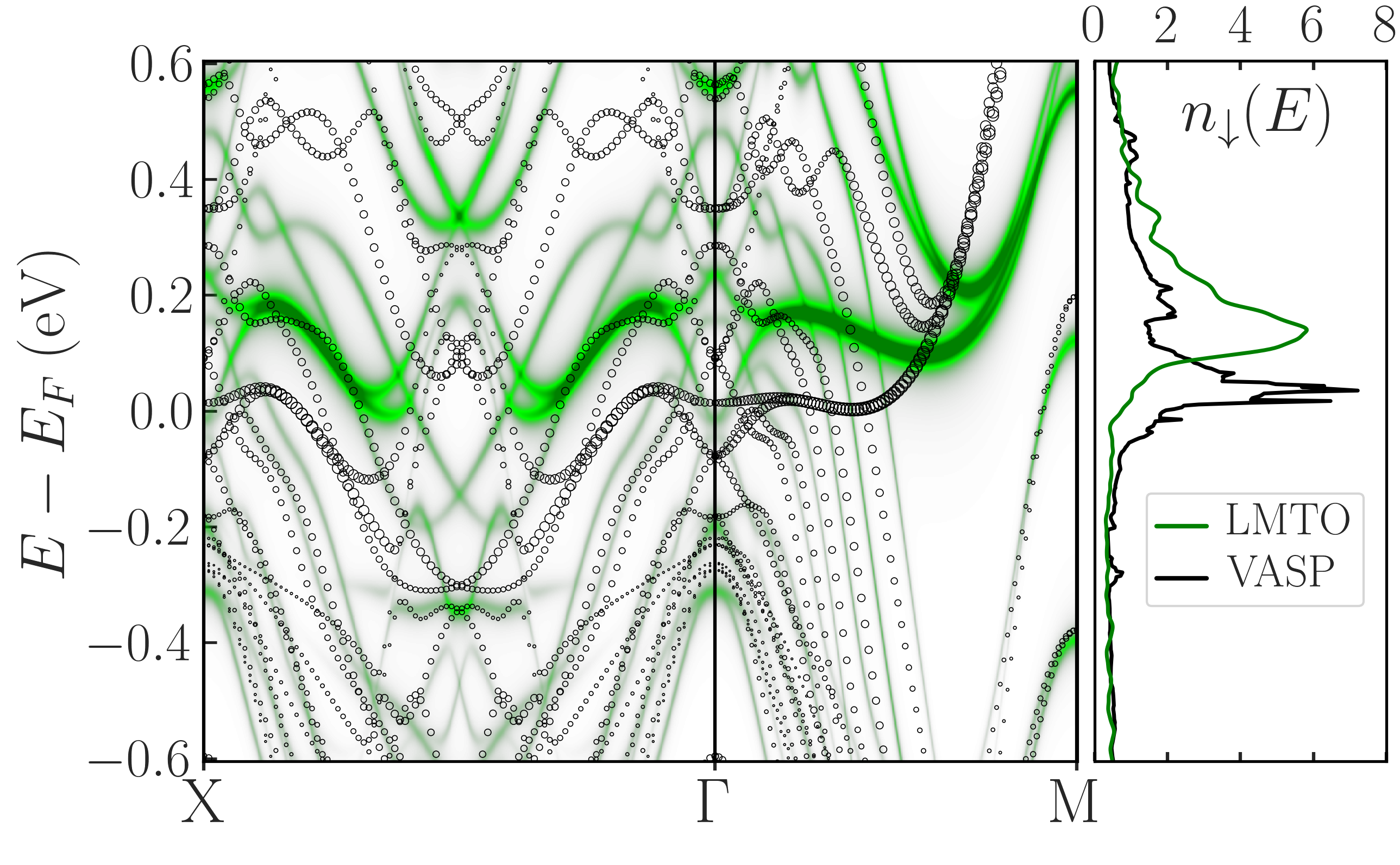}
    \caption{Minority-spin band structure for the Fe/MgO(4)/Fe MTJ highlighting the contribution of the interfacial Fe atoms. Black circles: VASP. Green density plot: LMTO with an empirical correction. Right panel: DOS (similar to Fig.\ \ref{fig:DOSALL}(a)).}
    \label{fig:BANDSDOS}
\end{figure}

Figure \ref{fig:BANDSDOS} shows the band structure from the empirically corrected LMTO calculation compared to the PBE calculation from VASP. It is seen that the dispersion of the interface bands is similar in both methods, and the empirical correction results in an upward shift of these bands.

\section{Tunneling in F\lowercase{e}/M\lowercase{g}O/F\lowercase{e} junctions}
\label{sec:femgofe}

In this section we study spin-dependent tunneling in Fe/MgO/Fe MTJs (Setup I) using the tight-binding linear muffin-tin orbital (TB-LMTO) method implemented in the Questaal code \cite{QUESTAAL}.

It is known that the calculated minority-spin contribution in the parallel (P) configuration of a thin symmetric MTJ can be overestimated due to the perfect matching of the interface states on both interfaces, which in practice is likely to be broken by disorder or applied bias \cite{KBMtj}. The effect of disorder can be estimated by calculating the spin-resolved conductances as a function of the imaginary part of energy. We found that an imaginary part of 5 mRy reduces the contribution of the minority-spin channel in the P configuration, in both corrected and uncorrected DFT, to less than 5\% for the MTJ with $N=3$, while at $N>3$ it becomes negligible. This confirms that the minority-spin contribution in the P configuration comes almost exclusively from the perfectly matched interface states. Therefore this contribution is discarded in all the following calculations of TMR.

Table \ref{tab:SIGMATMR} shows the calculated spin-resolved transmittances at the Fermi energy ($T_{\uparrow\uparrow}$, $T_{\downarrow\downarrow}$) in the parallel (P) configuration, excluding minority, and either spin channel ($T_{\uparrow\downarrow}$) in the antiparallel (AP) configuration, along with the TMR ratios defined as $R=(G_\mathrm{AP}-G_\mathrm{P})/G_\mathrm{AP}$ and the resistance-area products $RA$ in the parallel configuration for Fe|MgO($N$)|Fe MTJs. We see that the empirical correction, which shifts the interface states up and away from the Fermi level, results in a significant increase of the TMR, especially for MTJs with thin MgO barriers. This is explained by the reduced efficiency of symmetry-enforced spin filtering when the thickness of the barrier is small, which allows a considerable contribution of interface states to $G_\mathrm{AP}$. For example, at $N=3$ the TMR increases from 120\% to nearly 860\% when the correction is included. Because DFT underestimates the band gap (and hence the decay rates) in MgO, the upper limit for TMR in ultrathin Fe|MgO|Fe MTJs with $N=3$ may still be somewhat larger.

\begin{figure}[htb]
    \includegraphics[width=0.95\columnwidth]{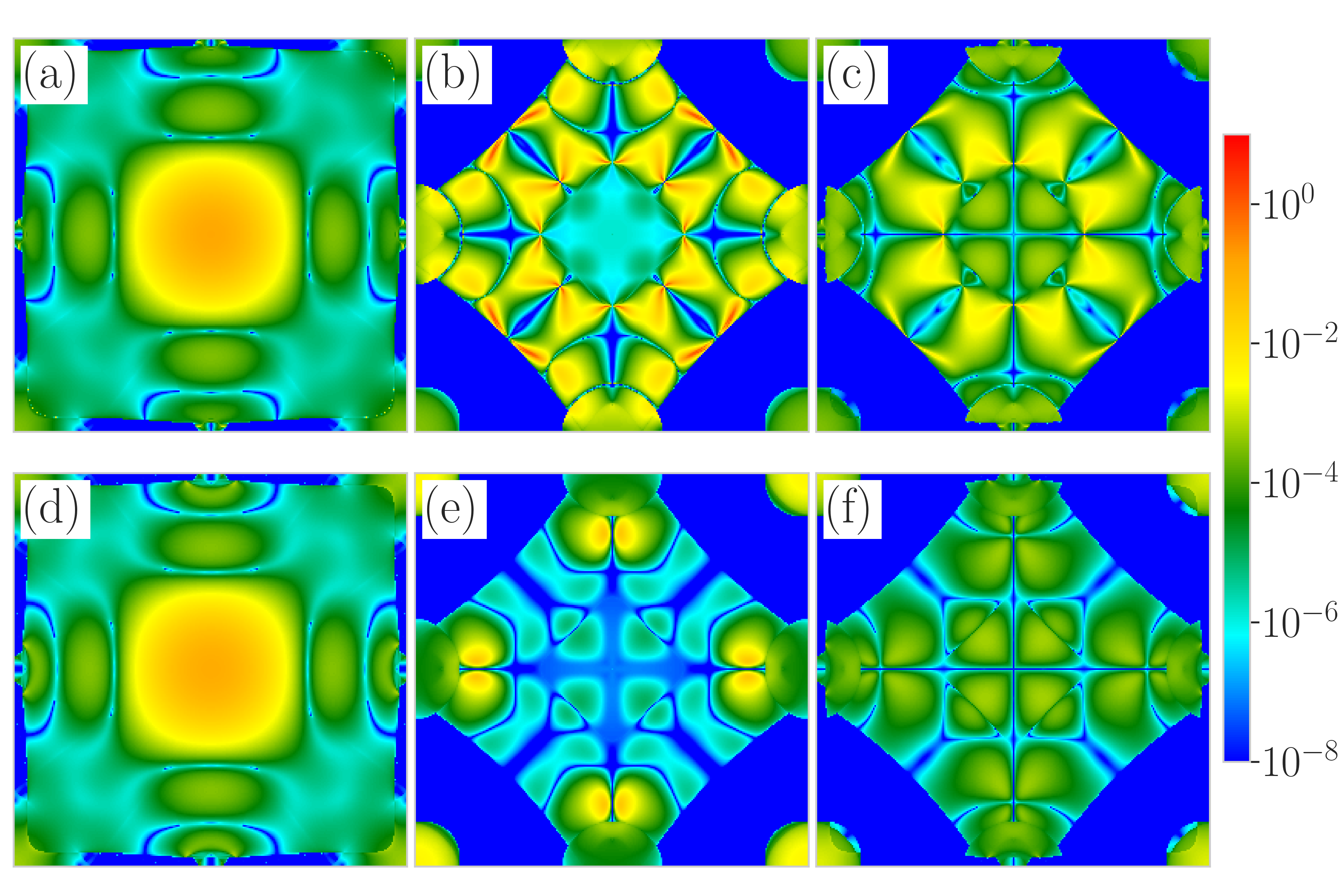}
    \caption{$\mathbf{k}_\parallel$-resolved zero-bias transmission functions in Fe|MgO(4)|Fe (logarithmic scale). (a)-(c) Uncorrected DFT. (d)-(f) DFT with an empirical correction for the interface states. (a), (d) $T_{\uparrow\uparrow}$. (b), (e): $T_{\downarrow\downarrow}$. (c), (f): $T_{\uparrow\downarrow}$.}
\label{fig:krtrans}
\end{figure}

\begin{table*}[htb]
        \caption{Spin-resolved conductance, TMR, and resistance-area product for Fe/MgO($N$)/Fe MTJ in the parallel ($T_{\uparrow\uparrow}$, $T_{\downarrow\downarrow}$) and antiparallel ($T_{\uparrow\downarrow}$) configurations (units of $e^2/h$ per unit cell), at the Fermi level. Columns marked as ``(corr)'' include calculations with the empirical correction for the interface states.}
            \begin{tabular}{|c|c|c|c|c|c|c|c|c|c|c|c|c|c|}
                \hline
                N & $T_{\uparrow\uparrow}$ & $T_{\downarrow\downarrow}$ & $T_{\uparrow\downarrow}$ & $T_{\uparrow\uparrow}$ (corr) & $T_{\downarrow\downarrow}$ (corr) & $T_{\uparrow\downarrow}$ (corr) & TMR & TMR (corr) & $RA_P$ (\unit{\ohm\micro\meter\tothe{2}}) & $RA_P$ (corr)\\
                \hline
                3 & 1.71$\times10^{-2}$ & 1.18$\times10^{-2}$ & 3.95$\times10^{-3}$ & 1.71$\times10^{-2}$ & 9.10$\times10^{-4}$ & 8.89$\times10^{-4}$ & 120\% & 860\% & 0.07 & 0.11 \\
                \hline
                4 & 4.00$\times10^{-3}$ & 4.84$\times10^{-3}$ & 3.41$\times10^{-4}$ & 3.63$\times10^{-3}$ & 1.67$\times10^{-4}$ & 5.86$\times10^{-5}$ & 490\% & 3000\% & 0.23 & 0.54 \\
                \hline
                6 & 1.98$\times10^{-4}$ & 6.99$\times10^{-6}$ & 3.24$\times10^{-6}$ & 1.82$\times10^{-4}$ & 9.82$\times10^{-7}$ & 1.83$\times10^{-6}$ & 2960\% & 4870\% & 10.1 & 11.3 \\
                \hline
                8 & 1.17$\times10^{-5}$ & 3.63$\times10^{-8}$ & 3.55$\times10^{-8}$ & 1.17$\times10^{-5}$ & 8.16$\times10^{-9}$ & 3.02$\times10^{-8}$ & 16400\% & 19300\% & 176 & 177 \\
\hline
            \end{tabular}
            \label{tab:SIGMATMR}
        \end{table*}
Figure \ref{fig:krtrans} shows the $k$-resolved transmission functions at the Fermi energy for both spin channels in Fe|MgO(4)|Fe in the P configuration and either spin channel in AP. The top row of panels (a-c) were calculated in DFT, and the bottom row (e-f) including the empirical correction for the interface states. As expected, the correction has no effect on the majority spins, cf. panels (a) and (d). However, the correction removes the ``hot spots'' coming from the interface states in $T_{\downarrow\downarrow}$ and $T_{\uparrow\downarrow}$, resulting in an enhanced TMR.

\begin{figure}[htb]
        \includegraphics[width=0.9\columnwidth]{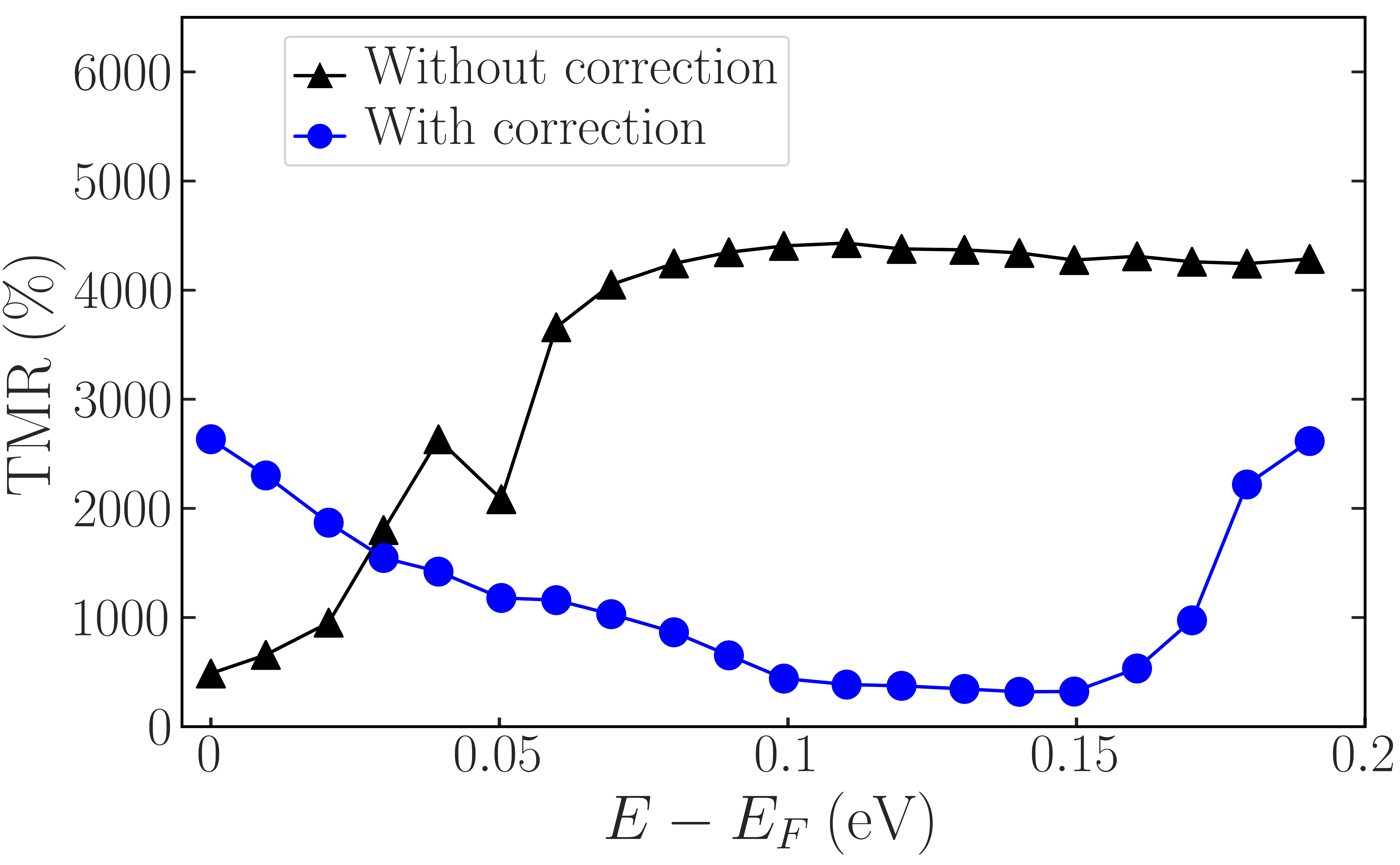}
        \caption{Zero-bias TMR for a Fe|MgO(4)|Fe MTJ as a function of the Fermi level shift with and without the empirical correction.}
        \label{fig:TMR}
\end{figure}

Figure \ref{fig:TMR} shows the dependence of the TMR on the Fermi energy for Fe|MgO(4)|Fe. In uncorrected DFT, the TMR increases with increasing Fermi energy until it reaches a plateau exceeding 4000\% at about 0.07 eV. With the correction, the TMR starts at a much higher value of about 3000\% and declines to well below 1000\% at 0.1 eV before sharply increasing again. This behavior reflects the effect of the interface states, which reduce TMR when they cross the Fermi energy not too far from the $\Gamma$ point. The minimum in the TMR corresponds to the situation when the Fermi energy passes through the flat region of the interface states along the $\Gamma M$ line, which is seen in Fig. \ref{fig:BANDSDOS}. These crossings appear in Fig. \ref{fig:krtrans} as ``hot spots'' along the diagonals of the Brillouin zone.

\section{Tunneling density of states and TMR in F\lowercase{e}X/M\lowercase{g}O/F\lowercase{e}X junctions}
\label{sec:TDOS}

We study the effects of alloying with Co or V using the coherent potential approximation (CPA). We consider FeX|MgO interfaces and FeX|MgO|FeX tunnel junctions, where FeX refers to a substitutional bcc Fe$_{1-x}$Co$_x$ or Fe$_{1-x}$V$_x$ alloy. Because Fe, Co, and V atoms are chemically similar, and the concentration of Co or V is not large, it is assumed that the structure of the junction is unaffected by alloying.

\begin{figure}[htb]
        \includegraphics[width=.95\columnwidth]{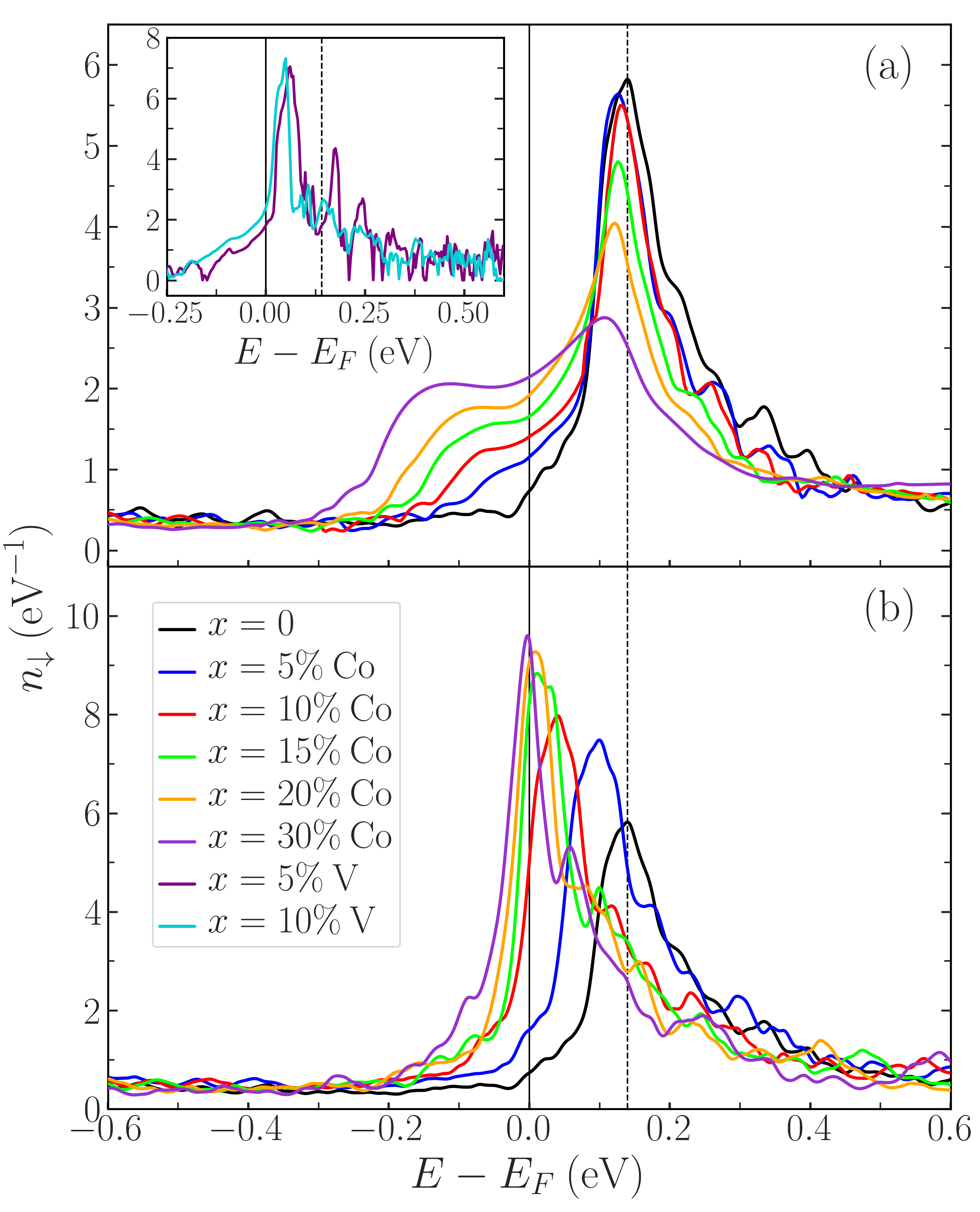}
        \caption{Partial DOS for the interfacial FeCo layer in a FeCo|MgO(4)|FeCo MTJ for different Co concentrations $x$ (see legends) calculated in (a) CPA, and (b) VCA. The empirical correction for the surface FeCo layer is included.}
        \label{fig:CPAVCADOS}
\end{figure}

\begin{figure}[htb]
        \includegraphics[width=0.95\columnwidth]{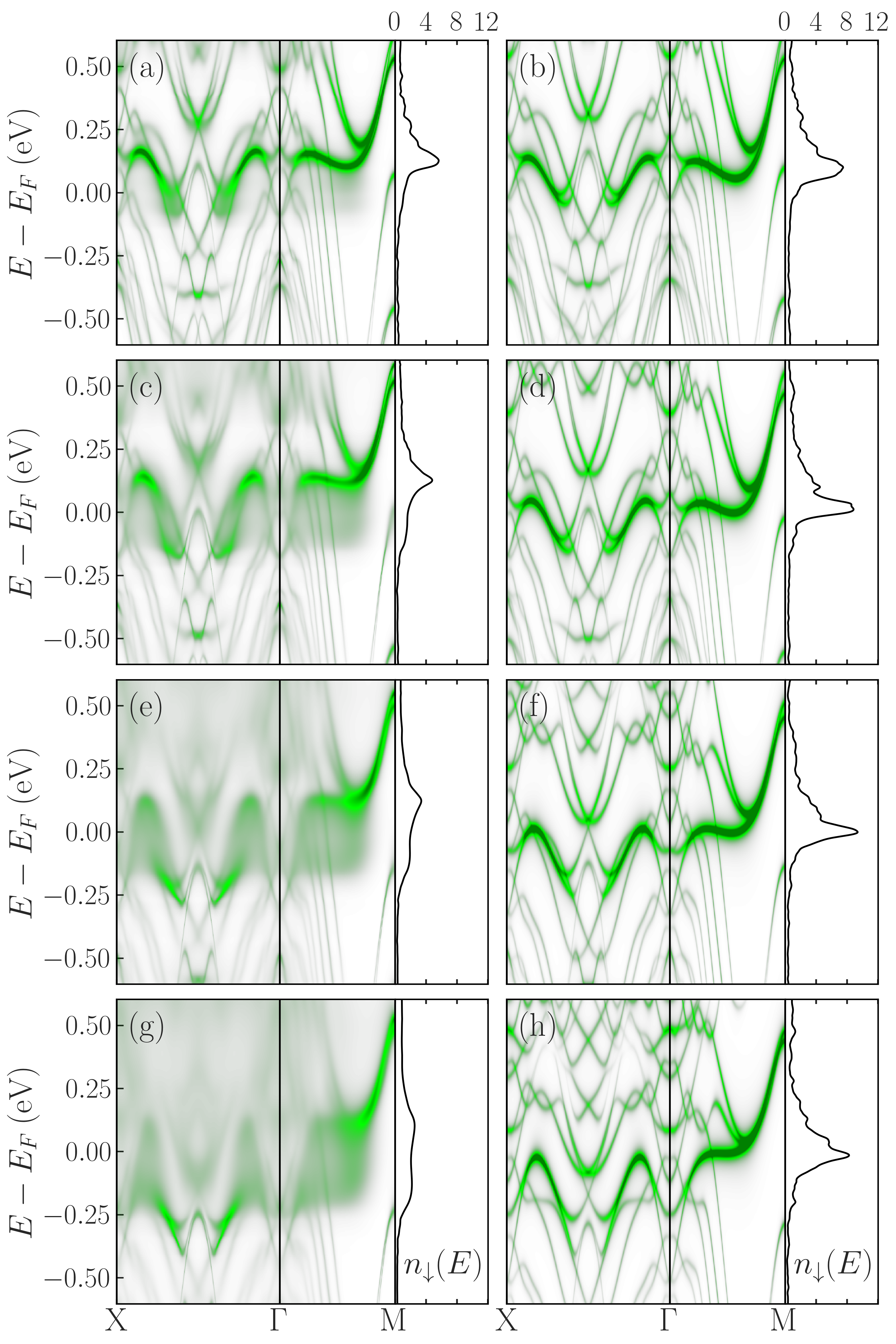}
        \caption{Spectral functions for the interfacial FeCo layer in FeCo|MgO(4)|FeCo MTJ calculated in CPA (left panels) or VCA (right panels). In VCA, a small imaginary part is added to the energy. (a-b) 5\% Co; (c-d) 15\% Co; (e-f) 25\% Co; (g-h) 35\% Co.}
        \label{fig:CPAVCASPF}
\end{figure}

In Fig.\ \ref{fig:CPAVCADOS}, the partial minority-spin DOS for the interfacial FeCo layer calculated in CPA is compared with the virtual crystal approximation (VCA) for a range of concentrations. Figure \ref{fig:CPAVCASPF} shows the corresponding partial Bloch spectral functions for that same layer. In VCA, alloying with Co shifts the interface states down, and they cross the Fermi energy at a rather small concentration of 10-15\% Co. On the other hand, in CPA the alloying results in a significant broadening of the interface states. This can be see both in the spectral functions in Fig.\ \ref{fig:CPAVCASPF} and in the partial DOS at the interfacial FeCo layer (Fig.\ \ref{fig:CPAVCADOS}) where the peak in CPA is almost unshifted while a broad shoulder appears at the Fermi energy. As seen in Fig.\ \ref{fig:CPAVCASPF}, the spectral weight corresponding to that shoulder is incoherent. On the other hand, alloying with V brings the minority-spin DOS peak much closer to the Fermi level, even though V has fewer electrons than Fe and in the rigid-band picture the Fermi level would move away from the peak. The rigid-band picture fails because V impurities in Fe have a large local moment antiparallel to that of the Fe host. CPA gives the V local moment of $-1.19 \mu_B$ in the bulk and $-1.57 \mu_B$ in the surface layer. Thus, alloying with V decreases the exchange splitting, counteracting the effect of hole doping.

\begin{figure}[htb]
        \includegraphics[width=0.9\columnwidth]{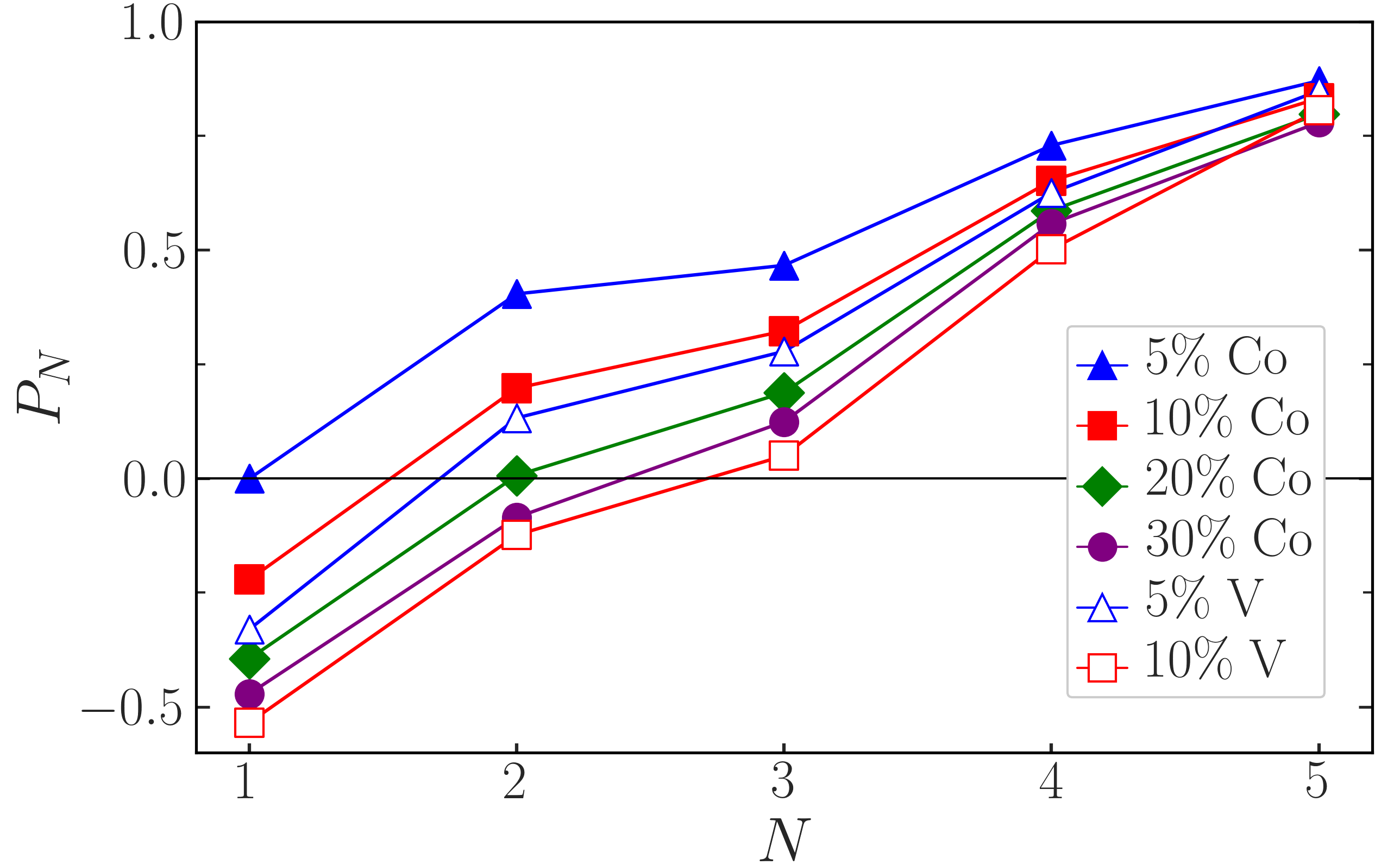}
        \caption{Spin polarization $P$ of the barrier DOS for MgO monolayers $N=1$ to 5, counting from the interface, in a Fe$_{1-x}$X$_{x}$/MgO(12)/Fe$_{1-x}$X$_{x}$ MTJ (X=Co,V).}
        \label{fig:SpinPol}
\end{figure}

Figure \ref{fig:SpinPol} shows the spin polarization of the $k$-integrated barrier DOS, defined as
\begin{equation}
P_N=\frac{n^{(N)}_\uparrow(E_F)-n^{(N)}_\downarrow(E_F)}{n^{(N)}_\uparrow(E_F)+n^{(N)}_\downarrow(E_F)}
\end{equation}
where $n^{(N)}_\sigma(E)$ is the partial DOS of spin $\sigma$ for the $N$-th MgO monolayer, counting from the interface. As expected, the spin polarization increases with increasing distance from the interface at all concentrations of Co. We also see that this spin polarization monotonically decreases with increasing concentration of Co.

If we were to use Julliere's formula to predict TMR from $P$ in the middle of the barrier, the results shown in Fig.\ \ref{fig:SpinPol} would indicate that TMR should quickly decline with increasing concentration of Co and vanish at $x\approx 0.3$ for 3-5 ML of MgO. However, a more refined prediction of TMR that takes symmetry-enforced spin filtering can be made on the basis of the $k$-resolved DOS. If the tunneling current at a given $\mathbf{k}_\parallel$ is dominated by a single evanescent barrier state, the transmission function of the MTJ, $T^\sigma(\mathbf{k}_\parallel)$, can be represented as a product of the surface transmission functions (STF) $T_{L,R}(\mathbf{k}_\parallel)$ for the left (L) and right (R) leads: $T^\sigma(\mathbf{k}_\parallel)=T_L^\sigma(\mathbf{k}_\parallel)T_R^\sigma(\mathbf{k}_\parallel)$ \cite{Belashchenko2004,TMR-theory}. $T_{L,R}^\sigma(\mathbf{k}_\parallel)$ is proportional to the DOS at the given $\mathbf{k}_\parallel$ induced in the barrier by that lead, at a distance corresponding to the middle of the barrier in the MTJ. This approximation is expected to become better with increasing barrier thickness, because tunneling states with larger decay parameters are exponentially suppressed. We will employ this approximation (which we will refer to as the STF method) to evaluate the TMR in FeX|MgO|FeX alloys with X=Co and V, treating substitutional disorder in CPA. Note that the configurational averaging of the product of two surface transmission functions does not involve any vertex corrections, because disorder in one lead has no effect on the surface transmission function of the other lead.

\begin{figure}[htb]
        \includegraphics[width=0.95\columnwidth]{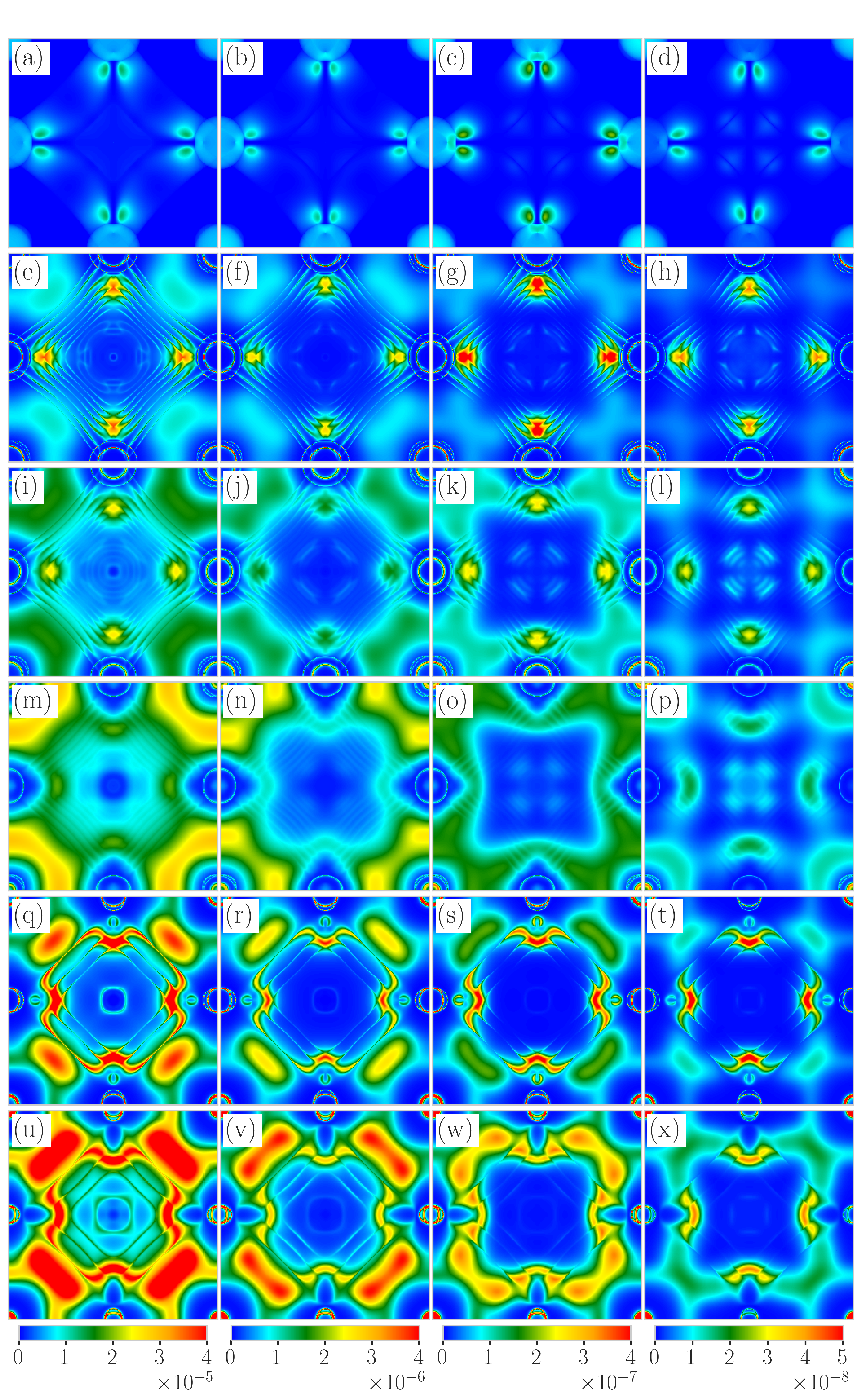}
        \caption{$\mathbf{k}_\parallel$-resolved minority-spin metal-induced partial DOS in MgO for X=Co,V concentration of 0\% Co (a-d), 5\% Co (e-h), 10\% Co(i-l), 20\% Co (m-p), 5\% V (q-t), and 10\% V (u-x) in FeX|MgO(12 ML)|FeX MTJs. Panels from left to right [e.g., (a) to (d)] correspond to consecutive MgO layers counting from the interface. At 0\% Co, DOS is calculated using the embedding method, otherwise a periodic supercell with 50 ML of the FeX(X=Co,V) alloy is used.}
        \label{fig:CPAMGO}
\end{figure}

We calculate $k$-resolved tunneling DOS using a periodic Fe(50)|MgO(12) system. A longer Fe layer helps reduce the effects of $k_z$ quantization leading to the formation of quantum-well states in Fe. In the majority-spin channel, quantum-well states are very pronounced regardless of the concentration of Co. Therefore, given that the addition of Co has a very small effect on the majority-spin states, we instead use the $k$-resolved DOS for pure Fe|MgO|Fe obtained in the embedding geometry. The minority-spin DOS is shown in Fig. \ref{fig:CPAMGO}, for three concentrations of Co and two of V, at the four layers of MgO near the interface. Here we see that substitutional disorder introduces sufficient broadening so that the quantum-well states are largely smeared out. Therefore, we can use the minority-spin $k$-resolved DOS obtained in CPA for the periodic system to obtain the TMR within the STF method.

From the results shown in Figs.\ \ref{fig:CPAVCASPF} and \ref{fig:SpinPol}, we expect an enhancement of the minority-spin spectral weight at the Fermi level with increasing concentration of Co. Figure \ref{fig:CPAMGO} shows that this enhancement comes predominantly at the periphery of the Brillouin zone, suggesting that TMR can remain sizeable after the addition of Co, especially at larger MgO thicknesses. The same argument applies to the addition of 5\% or 10\% V.

We now turn to the calculation of TMR using the STF method, i.e., using convolutions of $\mathbf{k}_\parallel$-resolved DOS in MgO. It is known that the calculated minority-spin contribution in the P configuration of a thin symmetric MTJs can be overestimated due to the perfect matching of the interface states on both interfaces, which in practice is likely to be broken by disorder or applied bias \cite{KBMtj}. Therefore, in the following we discard this minority-spin contribution.

\begin{figure}[htb]
        \includegraphics[width=0.85\columnwidth]{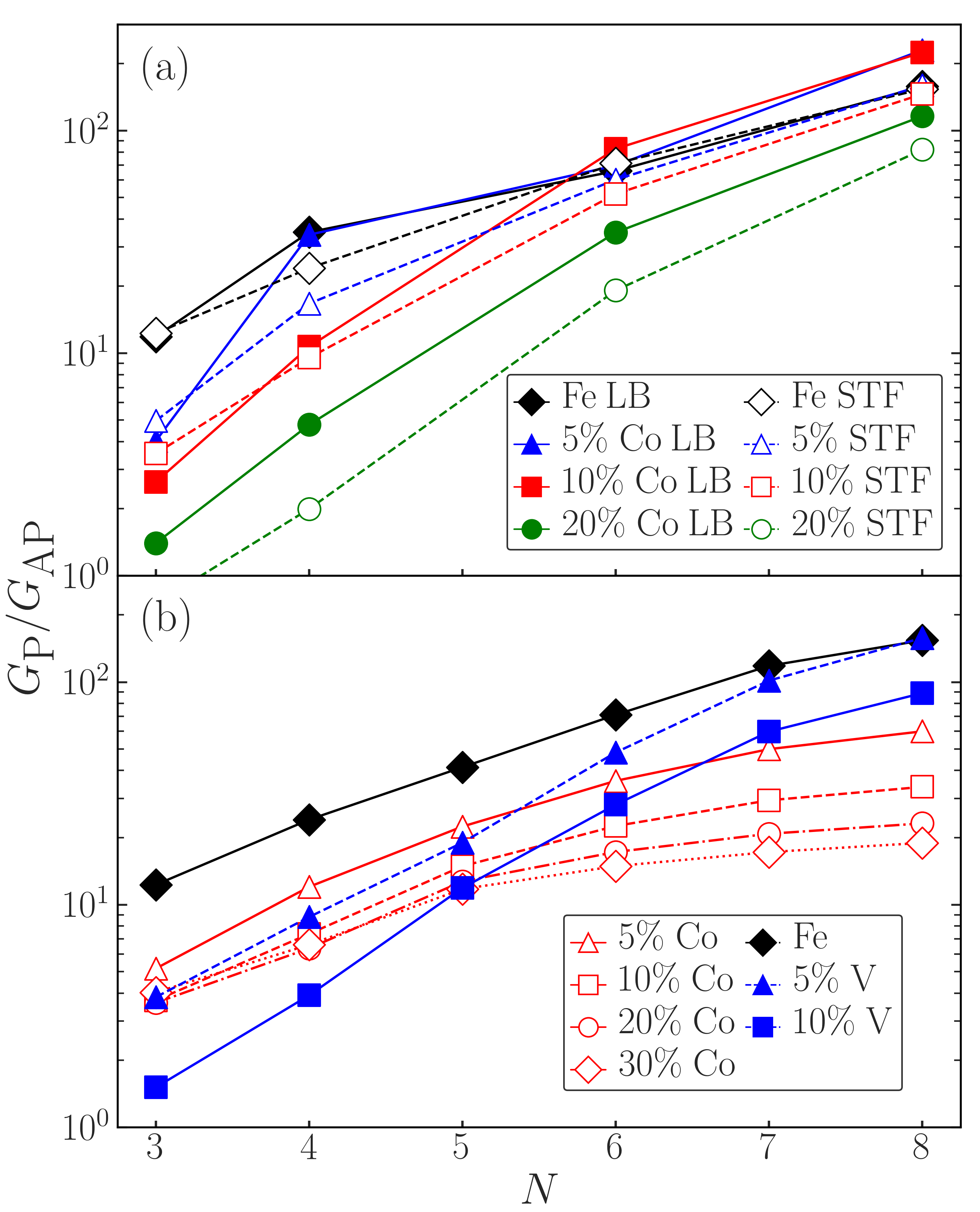}
        \caption{$G_\mathrm{P}/G_\mathrm{AP}$ ratio for (Fe-X)|MgO(N)|(Fe-X) MTJs (X=Co or V) as a function of MgO thickness $N$. The minority-spin channel is disregarded in the P configuration to rule out a spurious contribution from the matched interface states on both interfaces. (a) Comparison between the Landauer-B\"uttiker (LB) and surface transmission function (STF) methods within the virtual crystal approximation (VCA) for 0, 0.05, 0.1, and 0.2 extra electrons per Fe atom. (b) STF method with barrier DOS taken from CPA calculations with concentrations of Co or V indicated in the legend.
        }
        \label{fig:GCPA}
\end{figure}

To test the accuracy of the STF method we compare, in the virtual crystal approximation (VCA) for 0-20\% electron doping, the $G_\mathrm{P}/G_\mathrm{AP}$ ratio obtained using the Landauer-B\"uttiker and the STF methods. In this disorder-free case the DOS was necessarily obtained using the embedding method. For an odd number $N$ of MgO layers, we use DOS from MgO layer $(N+1)/2$. For an even number $N$, we calculate convolutions of DOS from layers $N/2$ and $N/2+1$. Figure \ref{fig:GCPA}(a) shows that the predictions of the two methods for the $G_\mathrm{P}/G_\mathrm{AP}$ ratio are in reasonable agreement with each other.

\begin{figure}[htb]
       \includegraphics[width=0.95\columnwidth]{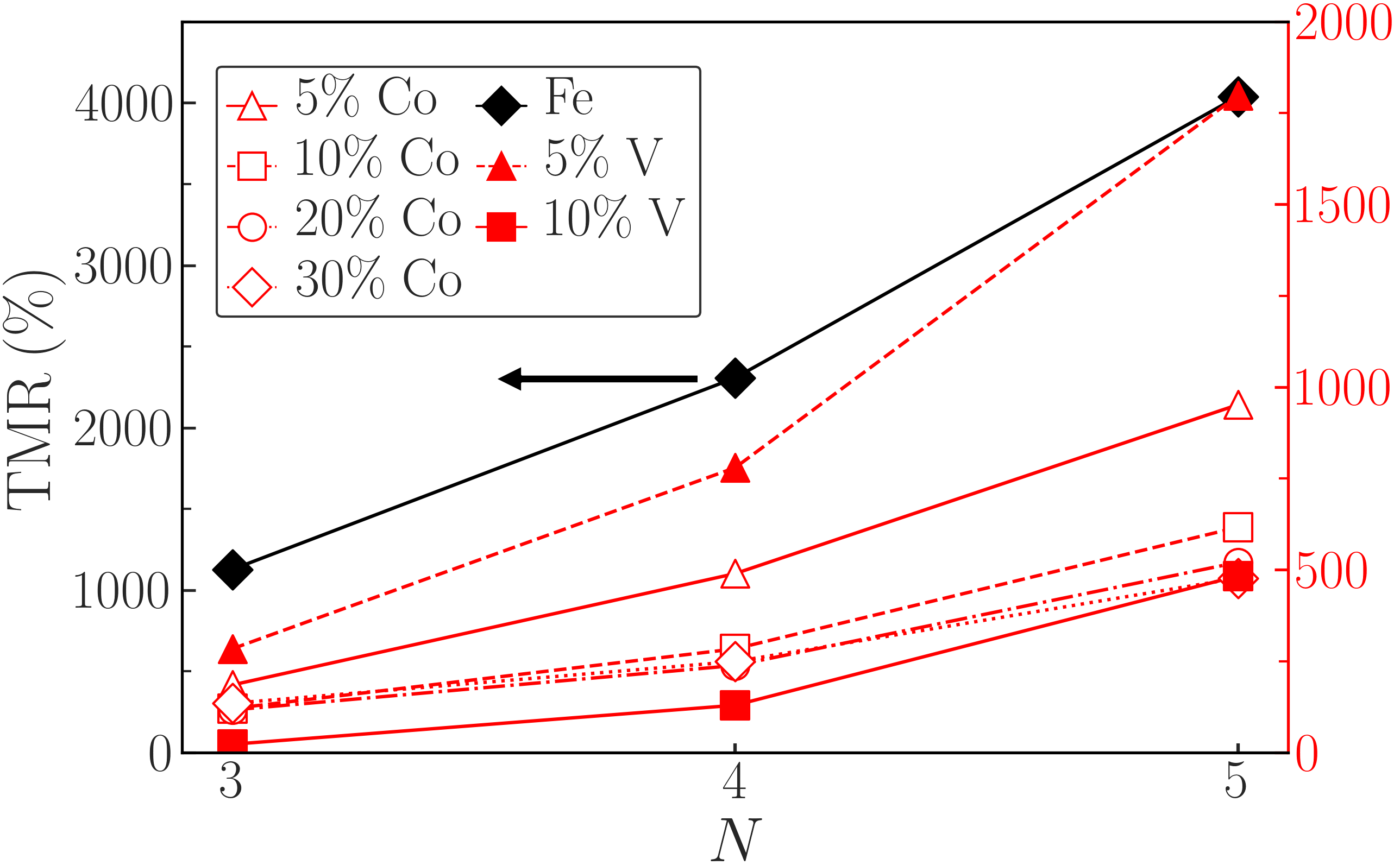}
        \caption{TMR ratio for (Fe-X)|MgO(N)|(Fe-X) MTJs (X=Co or V) in the 3-5 ML range of MgO thickness $N$.}
        \label{fig:TMRCPA}
\end{figure}

Figure \ref{fig:GCPA}(b) shows thickness-dependent $G_\mathrm{P}/G_\mathrm{AP}$ ratio calculated for FeCo|MgO|FeCo and FeV|MgO|FeV MTJs using the STF method with DOS taken from CPA (again, neglecting the minority-spin contribution in the P configuration), and Fig.\ \ref{fig:TMRCPA} shows the corresponding TMR in the 3-5 ML range. The $G_\mathrm{P}/G_\mathrm{AP}$ ratio and TMR are strongly suppressed by alloying compared to the case of pure Fe electrodes. This is due to the appearance of a large incoherent DOS at the Fermi level, as mentioned above in connection with Figs.\ \ref{fig:CPAVCASPF}-\ref{fig:CPAMGO}. However, the TMR is still on the order of 500\% at $N=4$ and increases to more than 1000\% at $N=5$, for both Co doping at any concentration and for V doping at 5\%. The persistence of relatively large TMR under alloying is due to the fact that the $\mathbf{k}_\parallel$-resolved minority-spin tunneling DOS is primarily concentrated at the periphery of the Brillouin zone (Fig.\ \ref{fig:CPAMGO}). Interestingly, after the addition of 10\% Co, the TMR ratio is not very sensitive to the Co concentration, especially at smaller MgO thicknesses.

\section{Bias dependence of TMR}
\label{sec:bias}

In this section, we examine the bias dependence of the TMR and the spin-resolved current densities in different spin channels. The current $I_{\sigma\sigma\prime}(V)$ flowing from spin channel $\sigma$ to spin channel $\sigma^\prime$ may be written, at zero temperature, as a convolution of the spin and $\mathbf{k}_\parallel$-resolved surface transmission functions $T^{\sigma}_{L/R}(\varepsilon,\mathbf{k}_{\parallel})$ \cite{KBMtj,TMR-theory} of the left (L) and right (R) leads, integrated over the bias window:
\begin{align}
  I_{\sigma\sigma\prime}(V) &= \sum_{\mathbf{k}_{\parallel}} \int\limits^{|e|V}_{0}T^{\sigma}_{L}(\varepsilon,\mathbf{k}_{\parallel})T^{\sigma^\prime}_{R}(\varepsilon-|e|V,\mathbf{k}_{\parallel}) d\varepsilon
  \label{current}
\end{align}
where the energy arguments of $T^{\sigma}_{L/R}(\varepsilon)$ are measured from the corresponding Fermi level, and we dropped the conductance quantum factor $e^2/h$. According to our convention, the current $I_{\sigma\sigma^\prime}(V)$ at positive bias $V$ is carried by electrons tunneling from filled states of spin $\sigma$ in the right lead into empty states of spin $\sigma^\prime$ in the left lead. As in Section \ref{sec:TDOS}, we use the tunneling DOS inside a 12-ML MgO barrier as an approximation for the surface transmission function. For the AP configuration with $N=4$, we use $T^{\uparrow}(\varepsilon,\mathbf{k}_{\parallel})$ and $T^{\downarrow}(\varepsilon,\mathbf{k}_{\parallel})$ evaluated at the second and third MgO layer, respectively.

\begin{figure}[htb]
       \includegraphics[width=0.95\columnwidth]{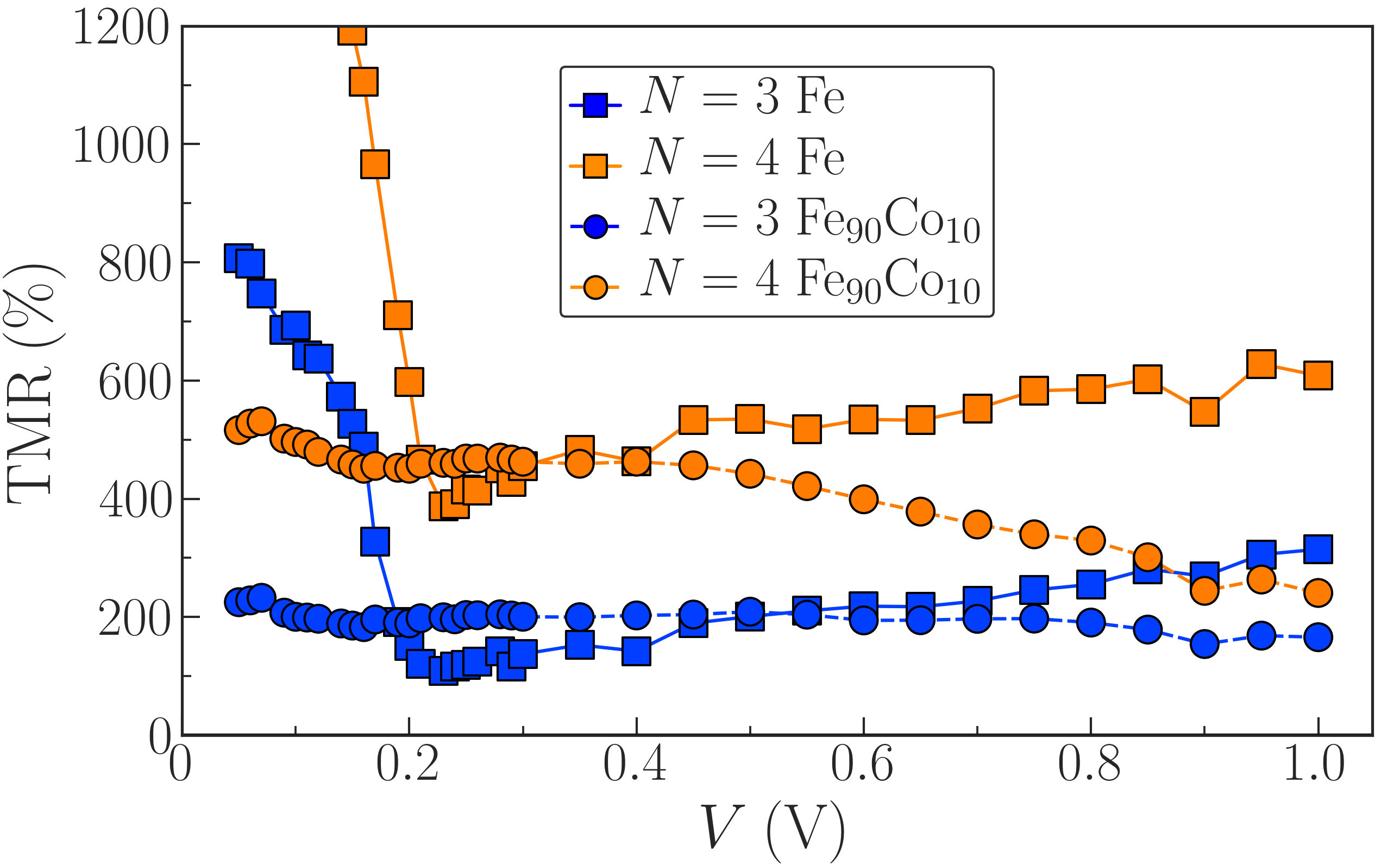}
        \caption{Bias dependence of TMR for MTJs with Fe and Fe$_{90}$Co$_{10}$ leads and $N=3$ or $4$ layers of MgO.}
        \label{fig:biasDependenceTMR}
\end{figure}

Figure \ref{fig:biasDependenceTMR} shows the TMR as a function of bias $V$ for FeX|MgO($N$)|FeX MTJs with pure Fe and Fe$_{0.9}$Co$_{0.1}$ leads with $N=3$ or 4. In the MTJs with Fe leads, the TMR decreases rapidly with increasing bias up to about 0.2 V and then slowly increases. In contrast, with Fe$_{0.9}$Co$_{0.1}$ leads the TMR is almost constant up to $V=0.5$ V. As already noted in Section \ref{sec:TDOS}, the TMR in MTJs with Co-doped leads is considerably reduced compared to those with pure Fe leads.

\begin{figure}[htb]
       \includegraphics[width=0.95\columnwidth]{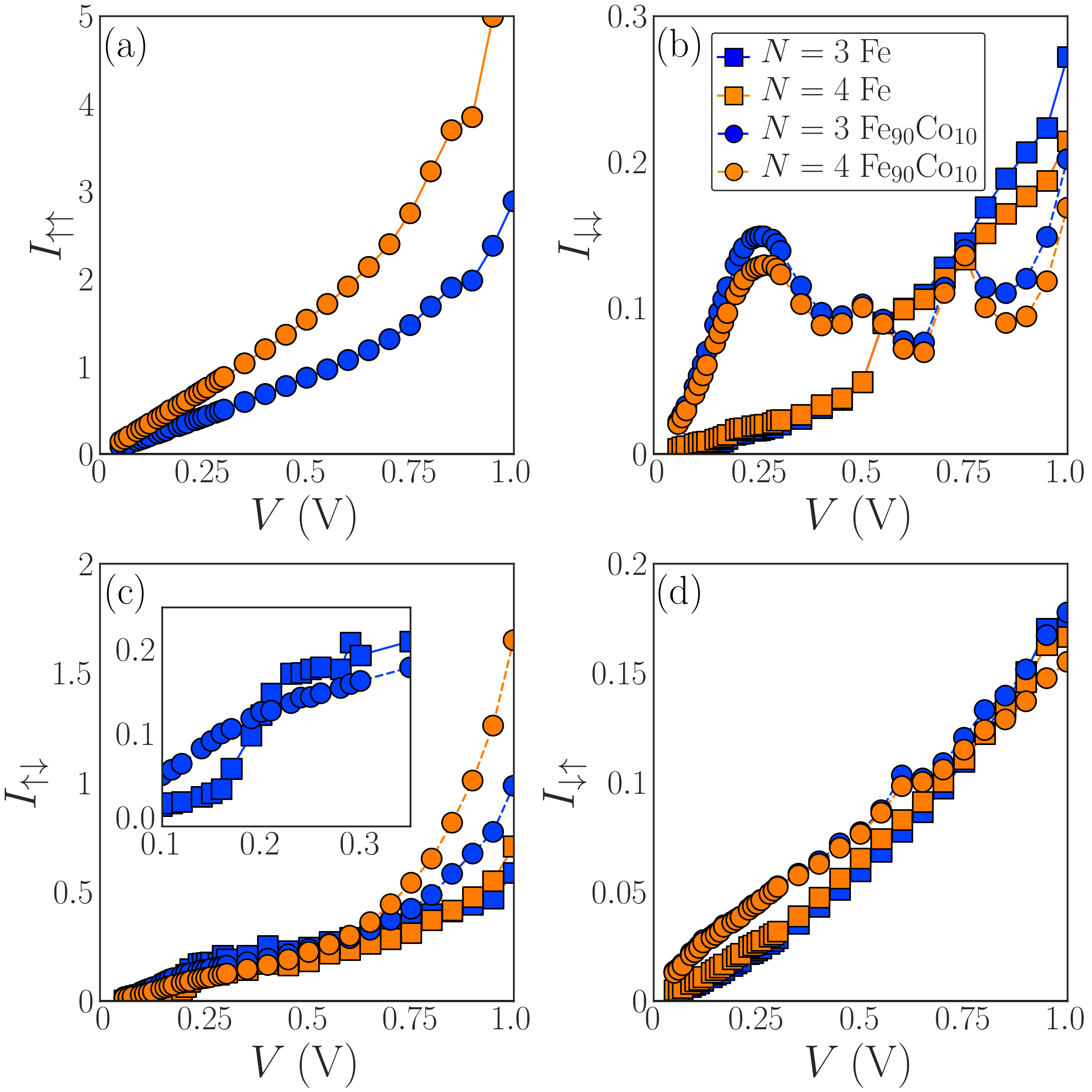}
        \caption{Bias dependence of the spin-resolved currents $I_{\sigma\sigma\prime}(V)$ for MTJ with Fe (squares) and Fe$_{90}$Co$_{10}$ (circles) leads. Blue (orange) symbols: 3 (4) ML of MgO. Data for $N=4$ are multiplied by 10.}
        \label{fig:biasDependence}
\end{figure}

To gain more insight into the drastically different bias dependence of TMR for Fe and Fe$_{0.9}$Co$_{0.1}$ leads, we examine the spin-dependent tunneling currents in the parallel and antiparallel configurations, shown in Fig. \ref{fig:biasDependence}. As expected, the current in the parallel configuration is dominated by the majority-spin contribution $I_{\uparrow\uparrow}(V)$, which increases linearly up to $V\approx0.5$ V \footnote{Because alloying with Co has almost no effect on the featureless majority-spin bands, we use the majority-spin $k_\parallel$-resolved DOS calculated with infinite leads for Fe/MgO/Fe. This allows us to avoid spurious quantum well states in the multilayer setup.}.
While the $I_{\downarrow\downarrow}(V)$ term is relatively small in both cases, it is strongly enhanced with Fe$_{0.9}$Co$_{0.1}$ leads due to the larger minority-spin spectral weight near the Fermi level, which was discussed in Section \ref{sec:TDOS} above. 
The nonmonotonic bias dependence of $I_{\downarrow\downarrow}(V)$ with Fe$_{0.9}$Co$_{0.1}$ leads occurs because at $V>$ \qty{0.3}{\volt} there are no filled states available at the $\mathbf{k}_\parallel$ points necessary to tunnel into the empty interface states.

In MTJ with Fe leads, the current in the antiparallel configuration is dominated by $I_{\uparrow\downarrow}(V)$, shown in Fig. \ref{fig:biasDependence}(c), where electrons from filled majority-spin bulk states tunnel into empty minority-spin interface states. This channel increases faster than linear as a function of $V$, leading to a sharp decline of  TMR that we noted in Fig. \ref{fig:biasDependenceTMR}. There is an especially fast increase in $I_{\uparrow\downarrow}(V)$ and a corresponding sharp drop in the TMR when the bias approaches \qty{0.2}{\volt}. In contrast, as seen in Fig. \ref{fig:biasDependence}, the bias dependence of all four channels in close to linear for MTJ with Fe$_{0.9}$Co$_{0.1}$ leads, which results in a nearly constant TMR.

\begin{figure}[htb]
       \includegraphics[width=0.85\columnwidth]{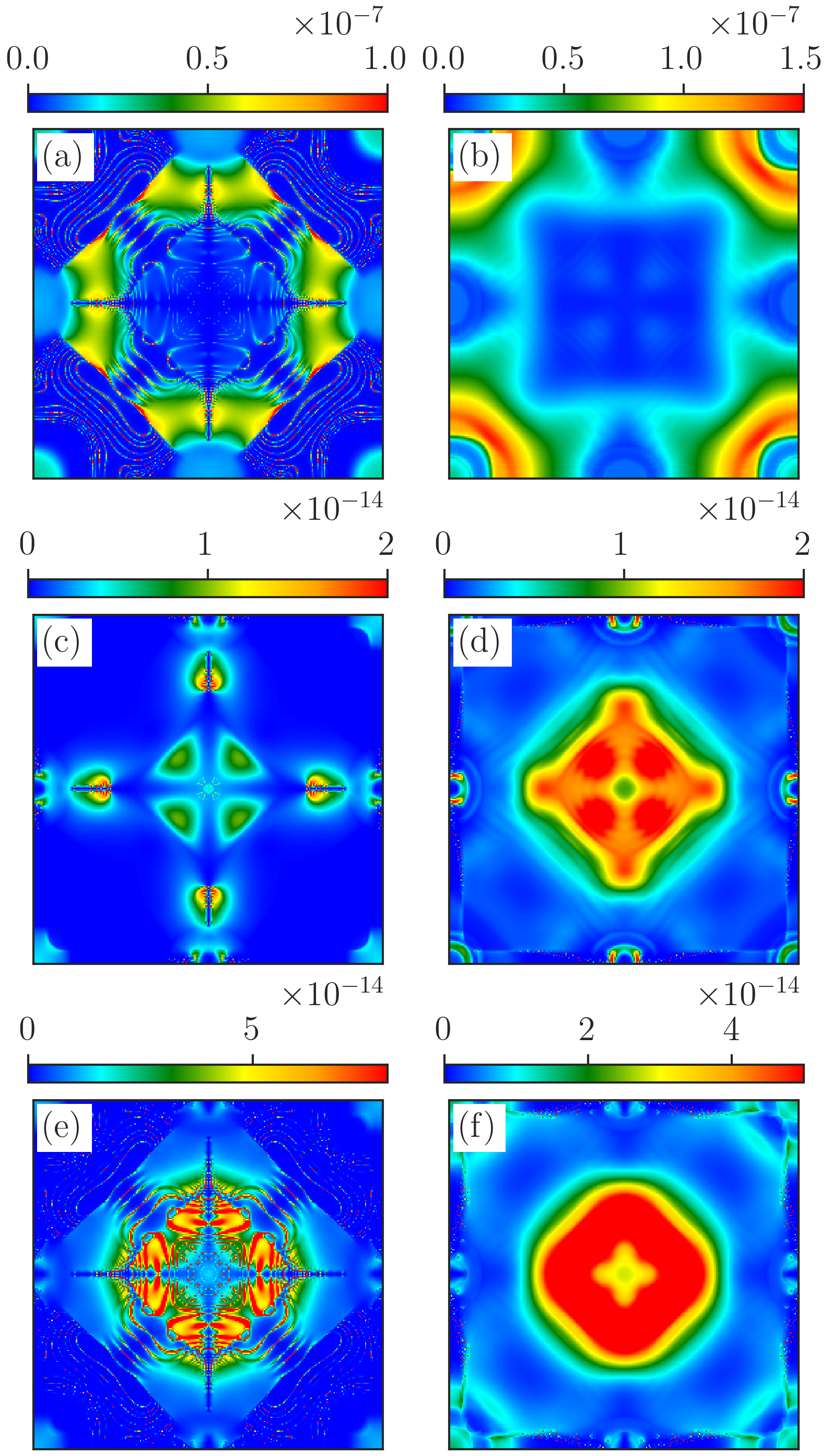}
        \caption{(a-b) $\mathbf{k}_\parallel$-resolved minority-spin DOS for the third layer of MgO in a 12-ML Fe/MgO/Fe MTJ, integrated from 0 to \qty{0.25}{\electronvolt} above $E_F$. (c-f) $\mathbf{k}_\parallel$-resolved currents $I_{\uparrow\downarrow}(V,\mathbf{k}_\parallel)$ at (c-d) $V=0.1$ \unit{\electronvolt} and (e-f) $V=0.25$ \unit{\electronvolt}. (a,c,d) Fe leads; (b,d,f) Fe$_{0.9}$Co$_{0.1}$ leads.}
        \label{fig:Tbar}
\end{figure}

Figure \ref{fig:Tbar}(a) shows the $\mathbf{k}_\parallel$-resolved minority-spin tunneling DOS for the third layer of MgO in the 12-ML Fe/MgO/Fe MTJ, integrated from 0 to \qty{0.25}{\electronvolt} above $E_F$. The interface states, seen as contours, are either localized or resonant depending on whether they fall in the bulk band gaps of the Fe leads at the given $\mathbf{k}_\parallel$. While the true integrated DOS is smooth, the discrete contours are seen because we used energy points spaced by 0.01 eV, which is sufficient for our purposes. Figure \ref{fig:Tbar}(b) shows the same quantity for the MTJ with Fe$_{0.9}$Co$_{0.1}$ leads. 

Figure \ref{fig:Tbar}(a) shows that the empty Fe/MgO interface states up to 0.25 eV above $E_F$ are spread over the entire Brillouin zone. In contrast, as seen in Fig. \ref{fig:Tbar}(b), the interface states at Fe$_{0.9}$Co$_{0.1}$/MgO are strongly broadened and located close to the corners of the Brillouin zone. This effect of alloying with Co is consistent with our earlier findings in Fig. \ref{fig:CPAMGO}.

Figures \ref{fig:Tbar}(c-d) show the $\mathbf{k}_\parallel$-resolved $I_{\uparrow\downarrow}(V,\mathbf{k}_\parallel)$ current channel at $V=0.1$ \unit{\electronvolt} for Fe and Fe$_{0.9}$Co$_{0.1}$ leads, respectively, and Figs. \ref{fig:Tbar}(e-f) show the same quantity at $V=0.25$ \unit{\electronvolt}. The drastic difference between Fig. \ref{fig:Tbar}(c) and Fig. \ref{fig:Tbar}(e) occurs because the interface states of the Fe leads, which are not too far from the $\bar\Gamma$ point, are brought into the bias window between $V=0.1$ and 0.25 \unit{\electronvolt}. This explains the large increase in $I_{\uparrow\downarrow}(V)$ seen in this bias range in Fig. \ref{fig:biasDependence}(c) and the abrupt drop of the TMR in Fig. \ref{fig:biasDependenceTMR}.

Because the partial DOS for the interfacial layer, seen in Fig. \ref{fig:CPAVCADOS}(a), is rather similar for Fe and Fe$_{0.9}$Co$_{0.1}$ leads, we might have expected a similar enhancement in $I_{\uparrow\downarrow}(V)$ and a drop in TMR in both cases when the peak in the partial DOS is brought inside the bias window. Figures \ref{fig:Tbar}(b), (d), and (f) show why this is not the case, and the TMR for Fe$_{0.9}$Co$_{0.1}$ leads is, instead, rather insensitive to bias. The reason is that the partial DOS peak for the Fe$_{0.9}$Co$_{0.1}$ leads is almost entirely composed of interface states in a narrow ring around the X point, which are effectively filtered out from the tunneling current. This also explains why TMR remains rather large in MTJs with Fe-Co leads, despite the appearance of a sizeable partial minority-spin DOS on the surface atoms. Still, the TMR is decreased compared to the case of pure Fe leads because some incoherent interfacial spectral weight does appear in the central part of the surface Brillouin zone. Figures \ref{fig:Tbar}(d) and (f) clearly show transmission assisted by this spectral weight. However, because it does not have a sharp structure as a function of energy, the resulting TMR is not sensitive to the bias voltage.

\section{Conclusions}
\label{sec:conclusions}

Motivated by the experimental evidence and GW results suggesting an incorrect placement of the minority-spin Fe/MgO interface states in DFT, we have introduced an empirical correction shifting these states up by about \qty{0.14}{\electronvolt} and studied its influence on the electronic structure and spin-dependent tunneling in FeX/MgO/FeX MTJs, where X stands for alloying with Co or V. As expected, we found that the correction strongly enhances the TMR in pure Fe/MgO/Fe junctions by removing the interface states from the Fermi level.

Alloying with Co is detrimental to zero-bias TMR at any concentration. By treating alloying via CPA, we find significant and consequential deviations from the rigid-band picture. Instead of gradually filling the interface states, alloying with Co broadens them and introduces a significant incoherent spectral weight at the Fermi level. However, this spectral weight is predominantly concentrated at the periphery of the Brillouin zone. As a result, the detrimental effect of Co is limited by spin filtering. For example, in MTJs with 4 ML of MgO, the TMR remains substantial at nearly 500\% and almost independent on the Co concentration in the range of 10-30\%. TMR of a similar magnitude is found for FeV|MgO|FeV MTJs with 5-10\% V.

Furthermore, the same shift of the spectral weight makes the TMR in MTJs with FeCo leads rather insensitive to bias up to about \qty{0.5}{\electronvolt}, in contrast to the MTJ with pure Fe leads where the TMR precipitously drops as the bias is increased from 0 to \qty{0.2}{\electronvolt}. This suggests that the bias dependence of TMR in FeCoB/MgO/FeCoB MTJs may, in practice, be dominated by impurity scattering or inelastic tunneling effects rather than by the intrinsic features of the electronic structure.

Experimentally, a moderate increase of TMR is often observed with the addition of Co \cite{Bonell}. If our assumption about the position of the interface states derived from Refs.\ \cite{Bonell,MTJGW} is correct, this increase is likely related not to the intrinsic electronic properties of the epitaxial interface but to extrinsic factors, such as the quality and homogeneity of the interface. On the other hand, the TMR predicted here for FeCo/MgO/FeCo MTJs is in reasonable agreement with the maximal reported experimental values \cite{Yuasa-handbook}.

\acknowledgments

We are grateful to Evgeny Tsymbal, Ilya Krivorotov, and Weigang Wang for their useful comments about the manuscript.
This work was supported by Seagate. M.v.S. was supported by National Renewable Energy Laboratory, operated by Alliance for Sustainable Energy, LLC, for the U.S. Department of Energy, Office of Science, Basic Energy Sciences, Division of Materials, under Contract No. DE-AC36-08GO28308. Calculations at UNL were performed utilizing the Holland Computing Center of the University of Nebraska, which receives support from the Nebraska Research Initiative. We also acknowledge the use of the National Energy Research Scientific Computing Center, under Contract No. DE-AC02-05CH11231 using NERSC award BES-ERCAP0021783, and also the Eagle facility located at the National Renewable Energy Laboratory.

\end{document}